\documentclass[aps,prl,twocolumn,showpacs,superscriptaddress,letter]{revtex4}
\usepackage{graphicx} 
\def\Journal#1#2#3#4{{#1} {\bf #2}, #3 (#4)}

\def\NIMA{{\rm Nucl. Instr. Meth.} A}
\def\NPA{{\rm Nucl. Phys.} A}
\def\PRL{\rm Phys. Rev. Lett.}
\def\PRC{{\rm Phys. Rev.} C}
 
\def\PL{\rm Phys. Lett.}
\def\ARNPS{\rm Ann. Rev. Nucl. Part. Sci.}
\begin{document}

\title{A Measurement of the Exclusive $^{3}\!He(e,e^{\prime}p)$ Reaction 
Below the Quasi-Elastic Peak}

\author{A.~Kozlov}
\affiliation{Department of Physics, University of Regina, Regina,
              SK S4S0A2, Canada}
\affiliation{School of Physics, The University of Melbourne, 
              VIC 3010, Australia}
\author{A.J.~Sarty}
\affiliation{Department of Astronomy \& Physics, Saint Mary's University,
              NS B3H3C3, Canada}
\author{K.A.~Aniol}
\affiliation{Department of Physics and Astronomy, California State
              University, Los Angeles, Los Angeles, CA 90032, USA}
\author{P.~Bartsch}
\author{D.~Baumann}
\affiliation{Institut f\"ur Kernphysik, Universit\"at Mainz,
              D-55099 Mainz, Germany}
\author{W.~Bertozzi}
\affiliation{Laboratory for Nuclear Science, 
              MIT, Cambridge, MA 02139, USA}
\author{K.~Bohinc}
\affiliation{Institute ``Jo\v zef Stefan'', 
              University of Ljubljana, SI-1001 Ljubljana, Slovenia}
\affiliation{Institut f\"ur Kernphysik, Universit\"at Mainz,
              D-55099 Mainz, Germany}
\author{R.~B\"{o}hm}
\affiliation{Institut f\"ur Kernphysik, Universit\"at Mainz,
              D-55099 Mainz, Germany}
\author{J.P.~Chen}
\affiliation{Thomas Jefferson National Accelerator Facility,
	      Newport News, VA 23606, USA} 
\author{D.~Dale}
\affiliation{Department of Physics and Astronomy,
              University of Kentucky, Lexington, KY 40506, USA}
\author{L.~Dennis}
\affiliation{Department of Physics,
              Florida State University, Tallahassee, FL 32306, USA}
\author{S.~Derber}
\author{M.~Ding}
\author{M.O.~Distler}
\affiliation{Institut f\"ur Kernphysik, Universit\"at Mainz,
              D-55099 Mainz, Germany}
\author{P.~Dragovitsch}
\affiliation{Department of Physics,
              Florida State University, Tallahassee, FL 32306, USA}
\author{I.~Ewald}
\author{K.G.~Fissum}
\affiliation{Laboratory for Nuclear Science, 
              MIT, Cambridge, MA 02139, USA}
\author{J.~Friedrich}
\author{J.M.~Friedrich}
\author{R.~Geiges}
\affiliation{Institut f\"ur Kernphysik, Universit\"at Mainz,
              D-55099 Mainz, Germany}
\author{S.~Gilad}
\affiliation{Laboratory for Nuclear Science, 
              MIT, Cambridge, MA 02139, USA}
\author{P.~Jennewein}
\author{M.~Kahrau}
\affiliation{Institut f\"ur Kernphysik, Universit\"at Mainz,
              D-55099 Mainz, Germany}
\author{M.~Kohl}
\affiliation{ Institut f\"ur Kernphysik, Technische Universit\"at
              Darmstadt, D-64289 Darmstadt, Germany}
\author{K.W.~Krygier}
\author{A.~Liesenfeld}
\affiliation{Institut f\"ur Kernphysik, Universit\"at Mainz,
              D-55099 Mainz, Germany}
\author{D.J.~Margaziotis}
\affiliation{Department of Physics and Astronomy, California State
              University, Los Angeles, Los Angeles, CA 90032, USA}
\author{H.~Merkel}
\author{P.~Merle}
\author{U.~M\"{u}ller}
\author{R.~Neuhausen}
\author{T.~Pospischil}
\affiliation{Institut f\"ur Kernphysik, Universit\"at Mainz,
              D-55099 Mainz, Germany}
\author{M.~Potokar}
\affiliation{Institute ``Jo\v zef Stefan'', 
              University of Ljubljana, SI-1001 Ljubljana, Slovenia}
\author{G.~Riccardi}
\author{R.~Roch\'{e}}
\affiliation{Department of Physics,
              Florida State University, Tallahassee, FL 32306, USA}
\author{G.~Rosner}
\affiliation{Department of Physics and Astronomy, University of Glasgow,
             Glasgow G12 8QQ, UK}
\affiliation{Institut f\"ur Kernphysik, Universit\"at Mainz,
              D-55099 Mainz, Germany}
\author{D.~Rowntree}
\affiliation{Laboratory for Nuclear Science, 
              MIT, Cambridge, MA 02139, USA}
\author{H.~Schmieden}
\affiliation{Institut f\"ur Kernphysik, Universit\"at Mainz,
              D-55099 Mainz, Germany}
\author{S.~ \v Sirca}
\affiliation{Institute ``Jo\v zef Stefan'', 
              University of Ljubljana, SI-1001 Ljubljana, Slovenia}
\author{J.A.~Templon}
\affiliation{Department of Physics and Astronomy,
              University of Georgia, Athens, GA 30602, USA}
\author{M.N.~Thompson}
\affiliation{School of Physics, The University of Melbourne, 
              VIC 3010, Australia}
\author{A.~Wagner}
\author{Th.~Walcher}
\author{M.~Weis}
\affiliation{Institut f\"ur Kernphysik, Universit\"at Mainz,
              D-55099 Mainz, Germany}
\author{J.~Zhao}
\author{Z.-L.~Zhou}
\affiliation{Laboratory for Nuclear Science, 
              MIT, Cambridge, MA 02139, USA}
\author{J.~Golak}
\affiliation{Institut f\"ur Theoretische Physik II,
Ruhr Universit\"at Bochum, D-44780 Bochum, Germany}
\affiliation{M. Smoluchowski Institute of Physics, Jagiellonian
University, PL-30059 Krak\'ow, Poland}
\author{W.~Gl\"ockle}
\affiliation{Institut f\"ur Theoretische Physik II,
Ruhr Universit\"at Bochum, D-44780 Bochum, Germany}
\author{H.~Wita\l{}a}
\affiliation{M. Smoluchowski Institute of Physics, Jagiellonian
University, PL-30059 Krak\'ow, Poland}
\collaboration{A1 Collaboration}\noaffiliation
\date{\today}

\begin{abstract}
New, high-precision measurements of the
$^{3}\!He(e,e^{\prime}p)$ reaction
using the A1 collaboration spectrometers at the Mainz microtron MAMI 
are presented. These were performed in antiparallel kinematics
at energy transfers below the quasi-elastic peak, and at a
central momentum transfer of 685 MeV/c.  Cross sections and
distorted momentum distributions were extracted
and compared to theoretical predictions and existing data. The
longitudinal and transverse behavior of the cross section was also studied.
Sizable differences in the cross-section behavior from theoretical predictions
based on Plane Wave Impulse Approximation were observed in both the
two- and three-body breakup channels.  Full Faddeev-type calculations 
account for some of the observed excess cross section, 
but significant differences remain.
\end{abstract}


\maketitle


Renewed precision studies of few-nucleon systems have been fueled
by recent developments on both the theoretical and experimental fronts.
Theoretically, progress toward full microscopic calculations based on realistic
$NN$ potentials has been achieved -- for example, via nonrelativistic
Faddeev-type calculations for three-body, \cite{Golak:95, Golak:01},
and via Monte Carlo variational calculations for few-body, systems
\cite{Carlson:94}. Experimentally, new facilities such as the Mainz microtron
MAMI and the Thomas Jefferson National Accelerator Facility provide 
high-quality continuous-wave electron beams, high-resolution magnetic 
spectrometers and fast data-acquisition. 
These features allow for precision measurements
of the nuclear electromagnetic response through coincidence proton-knockout 
electron scattering, and thus also for detailed studies of 
unresolved issues related to few-nucleon systems -- some
of which were first raised in inclusive $(e,e^{\prime})$ measurements 
more than 20 years ago.

The experiment reported in this paper is part of the systematic 
program ~\cite{proposal} 
to study nucleon-knockout reactions on $^{3,4}\!He$ nuclei
over the quasi-elastic (QE) peak at a fixed central
momentum-transfer of $| \vec{q} |$ = 685 MeV/c.
  The measurements reported here were performed
with $^{3}\!He$ on the low-energy-transfer side of the QE peak
(fixed central energy-transfer, $\omega \approx$ 158 MeV;
$x_B=(q^2 - \omega^2)/ 2M_p \omega \approx 1.5$),
where the contributions from isobars (IC) are suppressed, thereby 
enhancing the possibility of observing the effects of short-range $NN$
correlations.
%
%
\begin{figure}[t]
\includegraphics{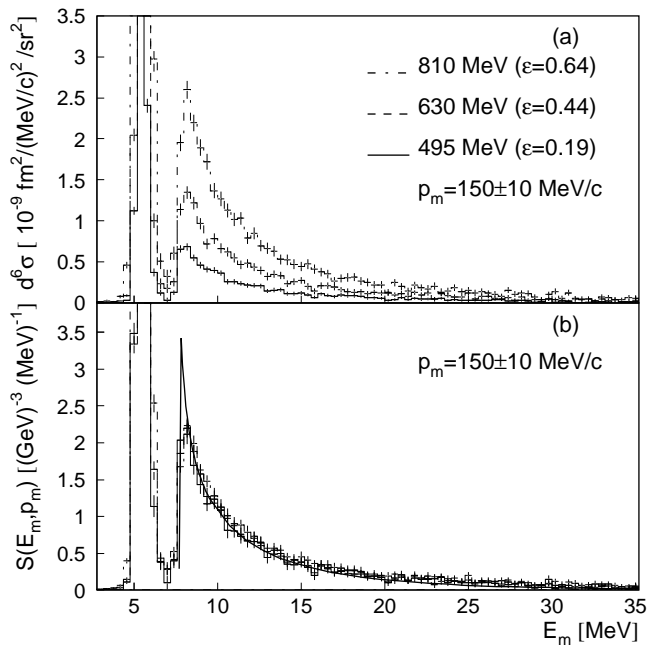}
\caption{\label{fig:xs} (a) Radiatively-corrected 
six-fold $^{3}\!He(e,e'p)X$ cross sections. The two-body breakup peak at 
5.49~MeV dominates the spectrum. The three-body breakup continuum
starts at 7.72~MeV.
(b) The experimental spectral function for the three measured
$\epsilon$-values compared to the theoretical calculations
from \cite{Kievsky:97}.}
\end{figure}
This low energy-transfer region was studied extensively in the 1970-80's
using inclusive electron scattering at $| \vec{q} | \gg k_F$
\cite{McCarthy:76, Day:78, Sick:80}.
The measured $^{3}\!He(e,e')$ cross
sections in this kinematical regime were much higher than theoretical
impulse approximation (PWIA)
predictions
\cite{Sick:80, degliAtti:83}, and  
the discrepancy with prediction increased as $\omega$ decreased (i.e. as
$x_B$ increased).  Since these inclusive cross sections were measured over
a wide range of momentum transfer and exhibited so-called ``$y$-scaling"
in this regime, it has been argued that the 
discrepancies cannot be accounted for by non-quasifree mechanisms
such as meson-exchange (MEC) or IC \cite{Day:90}.  Further, effects of final-state interactions
(FSI) in this region were estimated to be small \cite{Sauer:89}.
It was suggested \cite{Sick:80},
as a possible solution to this discrepancy,
that the high-momentum components in the $^{3}\!He$ spectral function should
be increased.  Such an increase would be indicative of short-range NN
correlations not accounted for in the theory.
This suggestion came under
scrutiny, and it was subsequently suggested \cite{degliAtti:83} that 
exclusive and semi-exclusive
$^{3}\!He(e,e^{\prime}p)$ measurements were required
to test such a modification, and to also examine the missing-energy ($E_{m}$)
dependence of the spectral function. In fact, a later measurement of
$^{3}\!He(e,e'p)$ at high missing momenta \cite{Marchand:88} ($p_{m}$) did
not support
{\it ad hoc} attempts to
increase the spectral function high-momentum components -- 
leading to the conclusion that the observed enhanced strength in 
the earlier inclusive data (as discussed in Ref. \cite{Sick:80})
was more likely due to defects in
the PWIA model assumptions.
However, it should be
noted that these exclusive data of Ref. \cite{Marchand:88} sampled
the kinematical domain with $x_B<$~1 (``dip"
region), and not the low-$\omega$ side at which the inclusive data are
discussed in Ref. \cite{Sick:80}.  In the low-$\omega$ region,
$y$-scaling was observed, indicating that single-nucleon reaction
mechanisms dominate.  Still, problems did exist with the calculations
and assumptions used in Ref. \cite{Sick:80}, since even at the QE peak
($x_B ~ \approx 1$), the theroetical calculations underpredicted that
magnitdue of the cross section data by as much as 20\%.
It can also be noted that none of the
since-reported exclusive $^{3}\!He(e,e'p)$ measurements
\cite{Jans:87, Ducret:93, LeGoff:97, Florizone:99} have been carried out
under the kinematical conditions studied by Ref.~\cite{Sick:80}.
In this work, we report the first
$^{3}\!He(e,e'p)$ measurements on the low-$\omega$
side of the QE peak.  These new exclusive data were taken with the 
momentum of the detected proton ($\vec{p_{p}}$) parallel to $\vec{q}$, making
the initial-proton momentum 
$\vec{p_{m}}=\vec{q}-\vec{p_{p}}$ antiparallel to $\vec{q}$
(``antiparallel kinematics").

The measurements utilized
three incident beam energies $E_{i}$~=~495, 630, and 810~MeV
with electron scattering angles $\Theta_e$~= 109.3$^\circ$, 
75.3$^\circ$, and 54.6$^\circ$, corresponding to virtual
photon polarizations of $\epsilon$~= 0.19, 0.44, and 0.64, where
$\epsilon =[1+2\frac{|\vec{q}|^{2}}{w^2 \,-\,q^2}\tan^{2}\frac{\Theta_e}{
2}]^{-1}$.
The total systematic uncertainty of the measured cross sections
was estimated to be 
$\pm3$\%. This uncertainty is dominated by the uncertainty in the
density of the $^{3}He$ gas in the cold ($T$ = 20 K), high-pressure
($P$ = 1823 kPa) target.
All error bars shown in the figures of this paper 
are statistical only. 
More details on the experimental setup, performance, and data-analysis 
can be found in Ref.\cite{Blomqvist:98} and \cite{Kozlov:99}. 

The $(e,e'p)$ cross section for $^{3}\!He$ was obtained as a
function of missing energy 
$E_{m}=\omega - T_p -T_{A-1}$ and $\vec{p_{m}}$, where $T_p$ and $T_{A-1}$ are kinetic energy of detected proton and recoil
$A-1$ system respectively.
The contribution from radiative processes were corrected in the 2-D
($E_m, p_m$) space using the Mainz version of the RADCOR code
\cite{Rokavec:94}; the results agree with those extracted using a second
method in which the radiative processes are accounted for in a Monte Carlo
simulation code ``MCEEP" \cite{mceep, Templon:00}.  It is found that the
observed strength beyond $E_m$~=~30~MeV is completely dominated by
radiative contributions and that, thus, the nuclear cross section (and
therefore the spectral function) becomes too small to be measured beyond
$E_m$~=~30~MeV.

The cross section in (anti)parallel kinematics can be
written as $\sigma \,=\,K (\sigma_T+\epsilon \sigma_L$), where K is a 
kinematical factor, $\sigma_T$ and $\sigma_L$ are the contributions to the 
cross section from  longitudinally ($L$) and transversely ($T$) polarized 
photons which can be separated by measurements at different $\epsilon$ (with
$\sigma$ here standing for the six-fold differential cross section).
Dividing $\sigma$ by the elementary $e-p$ cross section $\sigma_{ep}$ yields the
distorted spectral function:
\begin{equation} 
S^{dist}(E_{m},p_{m})=\frac{1}{p^{2}_{p}\sigma_{ep}^{cc1}}
\frac{d^{6}\sigma}{d\Omega_{e}d\Omega_{p} dp_{e}dp_{p}}
\end{equation}
%
%
\begin{figure}[t]
\includegraphics{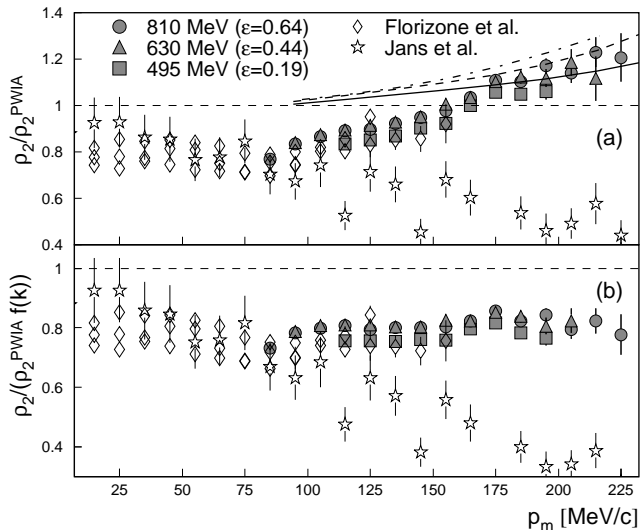}
 \caption{\label{fig:2bbu}
      {\it (a)}
      Ratio of extracted proton-momentum distributions for the 
      $^{3}\!He(e,e'p)d$ two-body breakup channel ($\rho_2^{exp}$)
      to the spectral function prediction ($\rho_2^{th}$) of Kievsky
      {\em et al.} \cite{Kievsky:97}.
      The data extracted from our measurements at the three values of
      $\epsilon$ are shown as full symbols. Earlier data from 
      MAMI ~\cite{Florizone:99}
      and Saclay ~\cite{Jans:87} are shown as open symbols.
      Note the Saclay data of Ref.~\cite{Jans:87} were taken under different
      kinematical conditions than the other data, being non-antiparallel and 
      fully centered at $x_B$ = 1.
      Theoretical curves ($\rho_2^{th} / \rho_2^{PWIA}$) ~\cite{Golak:95} shown are for 810~MeV beam energy 
      PWIA+FSI (solid),
      PWIA+FSI+MEC (dashed), and for 495~MeV beam PWIA+FSI+MEC (dashed-dotted);
      {\it (b)} Same ratio as in plot
      {\it (a)}, except now obtaining $\rho_2^{th}$ by multiplying the
      spectral function of Ref. ~\cite{Kievsky:97} by the renormalization
      factor $f(k)$ as suggested in Ref. ~\cite{Sick:80}.
      }
\end{figure}
Division of the cross section by a well-defined form of $\sigma_{ep}$
removes the predominant kinematic ($\epsilon$) dependence, enabling 
easier identification of residual nuclear dependencies. 
Physical interpretation of the data was done by comparison to: {\it (i)}
simple PWIA
based on the de Forest $cc1$ prescription \cite{Forest:83} for the elementary 
$e-p$ cross section ($\sigma_{ep}^{cc1}$), and {\it (ii)} full calculations 
using the Faddeev technique for the two-body breackup channel only
~\cite{Golak:95}. For the full calculations, the AV18 NN potential 
was used, and MEC were included; to isolate the effects of including
FSI and MEC on a PWIA starting point, comparison is made to the full 
calculations divided by PWIA -- both with MEC included (which we label
PWIA+FSI+MEC), and without MEC inclusion (labeled PWIA+FSI).
$\sigma_{ep}^{cc1}$ was selected for convenience in our PWIA description
to allow direct comparison to
earlier data from Ref.\cite{Florizone:99} and Ref. \cite{Jans:87}

The measured six-fold cross section for the reaction $^{3}\!He(e,e'p)X$
as a function of $E_m$, averaged over 
one sample $p_m$ bin of 150$\pm10$~MeV/c,
is shown for the three different $\epsilon$ values in Fig.~\ref{fig:xs}(a).
The dependence of the
cross section on the virtual-photon polarization is evident.
This dependence, however, disappears by dividing out the elementary
$ep$ cross section in the spectral function, as can be seen from Fig.~\ref{fig:xs}(b).
The residual $\epsilon$-dependence is less than 5~\% and shows no systematic 
trend, meaning that the $L/T$ behavior of the $(e,e'p)$ cross section is fully
described by the $\sigma_{ep}$.

Two-body $p-d$ (distorted) momentum distributions, $\rho_2$,
were obtained by
integrating the extracted $S^{dist}$ over the two-body breakup peak.  
The extracted $\rho_2$ distributions are shown in Fig.~\ref{fig:2bbu}(a)
plotted as a ratio with respect to the Kievsky spectral function. 
Like the spectral function itself, this ratio is independent of $\epsilon$ for
the entire $p_m$ range. Also shown in Fig.~\ref{fig:2bbu}(a) are earlier data
\cite{Florizone:99, Jans:87}. Several points are notable on the plot. First, 
a roughly constant 20\% suppression of the data 
compared to the spectral function is seen for $p_m$ up to 100 MeV/c.
This roughly 20\% difference between data and the Kievsky spectral function
at low $p_m$
has been previously observed (e.g. see Ref.~\cite{Florizone:99}) and has
been interpreted to be resulting 
predominantly from the fact that the theoretical spectral function does not 
account for FSI, and thus needs to be correspondingly renormalized.
Second,
agreement is seen between the current measurement and the earlier parallel
kinematics Mainz
data of Ref.~\cite{Florizone:99}, which were taken at the top of the QE peak.
Third, the data of Ref.~\cite{Jans:87} -- which were taken on top of the QE
peak, but is the only data-set taken in transverse kinematics,
thus at constant $\omega$ -- display reasonable agreement with those
of Ref.~\cite{Florizone:99} for $p_m$ up to 95 MeV/c, but deviate both from
our new data and those of Ref.~\cite{Florizone:99} at higher $p_m$.

The notable new feature
observed in Fig.~\ref{fig:2bbu}(a) is the significant relative 
increase in the measured cross section
compared to the PWIA for $p_m$ above 100 MeV/c.
The beginning of this phenomenon of relative increase
can be also seen in the few highest $p_m$ points of Ref.~\cite{Florizone:99}
but not in the data of Ref.~\cite{Jans:87}, which, in contrast to our new
data, show a steady decrease with respect to the calculation for 
increasing $p_m$.
It should be noted that the parallel kinematics data from
Ref.~\cite{Florizone:99} extend from the top of the QE peak to the
low-$\omega$ side, and that their $p_m$ dependence is 
coupled to the $\omega$ dependence from kinematical correlations, just as
for the new data.  This suggests that the (relative) increase in the
cross section is a phenomenon observed only at the
low-$\omega$ side.  Moreover, our
observed enhancement in strength appears to be consistent with that 
observed in the earlier inclusive measurements ~\cite{McCarthy:76, Day:78,Sick:80}. This can be demonstrated by multiplying
the spectral function of Kievsky {\em et al.} by the enhancement factor that
was suggested in Ref.~\cite{Sick:80}:
$f(k)=1+(k/\tilde{k})^n$
(where $n$=2.5, $\tilde{k}$=285~MeV/c, and $k$ is the primordial proton 
momentum in the nucleus, using $k$ = $p_m$ here).  When the theory is
modified in this way, then the $\rho_2^{exp}/\rho_2^{PWIA}$ ratio 
for our new data
becomes mainly constant at roughly 0.8 over the entire $p_m$ range up to
225 MeV/c. This is demonstrated in Fig.~\ref{fig:2bbu}(b).  

Finally, the last item to note with regard to Fig.~\ref{fig:2bbu}(a) is the
comparison to the full theory calculations.  The calculations including
effects of FSI and MEC merge to equal the PWIA calculation at the lowest
$p_m$ values ($p_m$ < 100 MeV/c); however, these full calculations still 
show a 20\% 
enhancement compared to data, even for the low $p_m$ region.  This overall
20\% difference may be due to the lack of consistent relativistic dynamics in 
the theory, due to the relatively high value of $q$.  Nevertheless, these
calculations provide the best current possibility of evaluating whether the
increase in the measured $\rho_2^{exp}/\rho_2^{PWIA}$ ratio, for $p_m$
above 100 MeV/c, is a manifestation of either FSI or MEC effects.
The full PWIA+FSI
curve shows an increasing trend similar to the data (increasing by 11\% as
$p_m$ increases from 100 MeV/c to 200 MeV/c),
and proper inclusion of MEC provides even more of a relative increase (with
an 17\% increase between $p_m$ = 100 and 200 MeV/c) --  but the data
still exhibit a greater still relative increase: 35\% between
$p_m$ = 100 and 200 MeV/c.
This result suggests
that FSI and MEC neglected by Ref.~\cite{Sick:80} are important and may explain
a substantial part of the discrepancy between data and PWIA.  Thus, it appears
the $f(k)$ enhancement factor is largely accounting for FSI and MEC -- however,
it should be noted again that the full calculations still do not show as a rapid
an increase (as a function of $p_m$) as do our new data.
Finally, the full calculations also suggest
that $\epsilon$-dependence is small ($<$~5\%) and comparable with our
systematic uncertainty, as shown on  Fig.~\ref{fig:2bbu}(a) for the 495 and 810~MeV
beam energies.
%
%
\begin{figure}[t]
\includegraphics{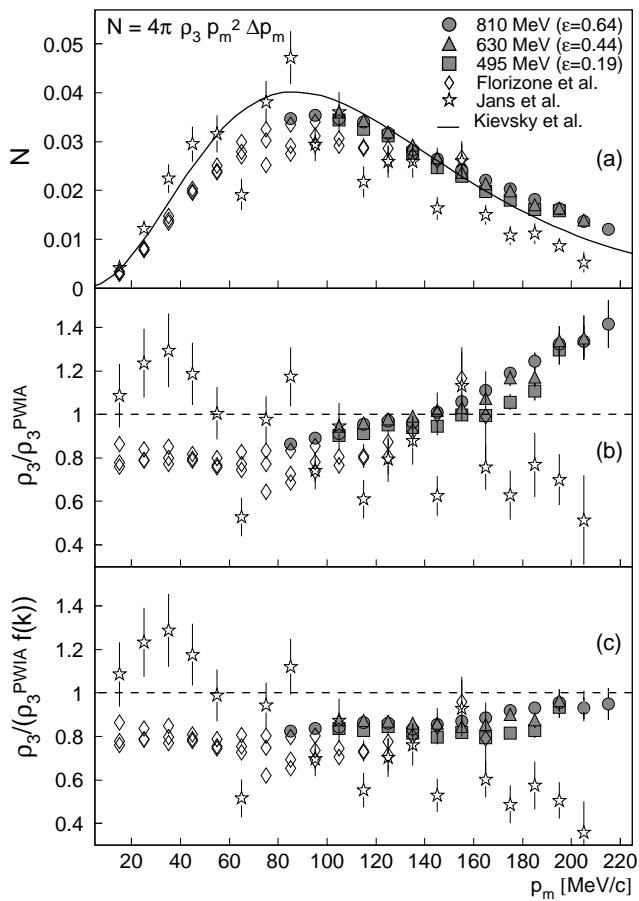}
 \caption{\label{fig:3bbu}
      Distorted proton-momentum distributions, $\rho_3$,
      for the $^{3}\!He(e,e'p)pn$ three-body breakup channel.  Plots and
      symbols follow the identical format used for
      $\rho_2$ in Fig.~\ref{fig:2bbu}.
      }
\end{figure}

Distorted proton-momentum distributions for the three-body
breakup channel, $\rho_3$, were obtained by integration of the
experimental spectral function $S^{dist}(E_{m},p_{m})$ from 7 to
20~MeV in $E_m$.  These extracted $\rho_3$ distributions are shown in
Fig.~\ref{fig:3bbu}(a), mutliplied by $4 \pi p_m^2 \Delta p_{m}$.
Again, no systematic $\epsilon$-dependence is observed.  The
ratio $\rho_3^{exp}$/$\rho_3^{PWIA}$ is shown as a function of $p_m$ in
Fig.~\ref{fig:3bbu}(b) using the unaltered theoretical prediction of
Kievsky {\em et al.}.  These ratios for the continuum
channel show a similar feature as seen in the two-body channel.
In Fig.~\ref{fig:3bbu}(c),
the prediction as modified by the enhancement
factor $f(k)$ from Ref.~\cite{Sick:80} is shown, leading to a better
agreement with the data, but not as good as was observed for the two-body
breakup channel.
  
In conclusion, we have performed measurements of the $^{3}\!He(e,e'p)$
reaction on the low $\omega$ side of the QE peak.
Cross sections, distorted spectral functions and distorted proton-momentum
distributions were obtained as a function of $E_m$ and 
$p_m$ for both the two-body and continuum breakup channel
for three $\epsilon$ values. The $(e,e'p)$ cross
section strength falls to near zero for $E_m~>$~30~MeV.
The strong $\epsilon$-dependence of the
cross section, both for two- and three-body reaction channels,
is due to the $L/T$ behavior of the elementary $e-p$ cross section.
The distorted proton-momentum distributions for 
both the two- and three-body breakup channels significantly
deviate in shape from the PWIA predictions of Ref.~\cite{Kievsky:97}
for $p_m$ above 100 MeV/c. Full calculations using the Faddeev technique 
show a roughly 20\% over-prediction at low $p_m$, but also show a functional
dependence on $p_m$, above 100 MeV/c, that compares to the data
significantly better than does PWIA -- indicating that the 
contributions from FSI and MEC neglected in earlier PWIA analyses
(e.g. ~\cite{Sick:80}) are not small and should be accounted for. 
Thus, although the new generation of theoretical calculations have changed 
the normalization issue, the physics issues surrounding high $p_m$
remain unresolved -- in fact, 
even the full calculations do not give satisfactory description of the data
at $p_m$ above 100 MeV/c, which could
be indicative of relativistic effects in the dynamics.
The correction function suggested in Ref.~\cite{Sick:80} to enhance the 
high-momentum components of the spectral function, based on
inclusive measurements in this same kinematical region ($\omega <
\omega_{QE}$) is therefore seen to account for deviations from PWIA 
in $p_m$ dependence arising
predominantly from FSI and MEC. Addtional work is needed, both experimental 
and theoretical, to identify the source of the remaining discrepancy between
our new data and the full calculations.

We would like to thank the MAMI staff
for providing support for this experiment.
This research was supported by the state of Rhineland-Palatinate and
by grants from the Deutsche Forschungsgemeinschaft (SFB 443),
by the U.S. Department
of Energy and the National Science Foundation, by the Natural Sciences
and Engineering Research Council, by Nato, by the Polish Committee for
Scientific Research, and by the University of Melbourne.  Faddeev 
calculations were performed on the Cray SV1 and the T3E of the NIC,
Juelich, Germany.


\begin{thebibliography}{10}

\bibitem{Golak:95} J.~Golak \emph{et~al.}, \Journal{\PRC}{51}{1638}{1995}.

\bibitem{Golak:01} J.~Golak \emph{et~al.}, \Journal{\PRC}{63}{034006}{2001}.

\bibitem{Carlson:94} J.~Carlson, and R.~Schiavilla,
\Journal{\PRC}{49} {2880} {1994}.

\bibitem{proposal} MAMI Experiments A1/3-96
and A1/1-93,
Institut f\"{u}r Kernphysik, Mainz University (R. Neuhausen and S. Gilad,
spokespersons).

\bibitem{McCarthy:76} J.S. McCarthy {\it et al.},
\Journal{\PRC}{13} {712} {1976}.

\bibitem{Day:78} D. Day {\it et al.},
\Journal{\PRL}{43} {1143} {1978}.

\bibitem{Sick:80} I.~Sick, D.~Day, and J.S.~McCarthy,
\Journal{\PRL}{45} {871} {1980}.

\bibitem{degliAtti:83} C.~Ciofi~degli~Atti, E.~Pace, and G.~Salme,
\Journal{\PL}{127B} {303} {1983}.

\bibitem{Day:90} D.B.~Day {\it et al.},
\Journal{\ARNPS}{40} {357} {1990}.

\bibitem{Sauer:89} H.~Meier-Hadjuk, and P.U.~Sauer,
\Journal{\NPA}{499} {669} {1989}.


\bibitem{Marchand:88} C.~Marchand \emph{et~al.}, 
\Journal{\PRL}{60} {1703} {1988}.

\bibitem{Florizone:99} R.E.J.~Florizone \emph{et~al.},
\Journal{\PRL}{83} {2308} {1999}.

\bibitem{Jans:87} E.~Jans \emph{et~al.}, \Journal{\NPA}{475} {687} {1987}.

\bibitem{Ducret:93} J.E.~Ducret \emph{et~al.},
\Journal{\NPA}{A553} {697} {1993}.

\bibitem{LeGoff:97} J.M.~Le~Goff \emph{et~al.}, \Journal{\PRC}{55}
{1600} {1997}.

\bibitem{Blomqvist:98} K.I. Blomqvist \emph{et~al.},
\Journal{\NIMA}{403} {263} {1998}.

\bibitem{Kozlov:99} A.~Kozlov, PhD thesis, Melbourne University (1999).

\bibitem{Rokavec:94} A.~Rokavec, 
PhD thesis, University of Ljubljana (1994).

\bibitem{mceep} P.E.~Ulmer, JLab Tech. Note, {\bf TN91-101} (1991).

\bibitem{Templon:00} J.A.~Templon {\it et al.},
\Journal{\PRC}{61} {014607} {2000}.

\bibitem{Forest:83} T.~de Forest \emph{et~al.},
\Journal{\NPA}{392} {232} {1983}.

\bibitem{Kievsky:97} A. Kievsky {\it et al.},
\Journal{\PRC}{56} {64} {1997}.


\end{thebibliography}
\end{document}